\documentclass[aps,prl,twocolumn,groupedaddress]{revtex4}

\usepackage{graphicx}
\usepackage{amssymb}

\newcommand{\rs}{\rm \scriptscriptstyle}

\newcommand{\sgn}{\mathop{\mathrm{sgn}}}

\begin{document}

\title {Crystalline phase for one-dimensional ultra-cold atomic bosons}

\author{Hans Peter B\"uchler}
\affiliation{Institute for Theoretical Physics III, University of Stuttgart, Germany}

\date{\today}

\begin{abstract}
We study cold atomic gases  with a contact interaction and confined into one-dimension. 
Crossing the confinement induced resonance  the correlation between the bosons increases, and 
introduces an effective range for the interaction potential. Using the mapping onto the sine-Gordon model 
and a Hubbard  model in the strongly interacting regime allows us to derive the phase diagram in the presence of an optical lattice. 
We demonstrate the appearance  of a phase transition from a Luttinger liquid with algebraic correlations into  a  crystalline phase  with a particle on every second lattice site.
\end{abstract}


\maketitle

Cold atomic gases confined into one-dimension exhibit remarkable properties as the 
interplay between interactions and reduced dimensions strongly enhances 
quantum fluctuations.
The most prominent example is the appearance of a Tonks-Girardeau gas for
bosonic particles \cite{kinoshita04,girardeau60}, and the possiblity to pin the bosons into a Mott insulating phase for arbitrary 
weak optical lattices \cite{haller10,buechler03}. Most remarkably, it has recently been proposed  \cite{astrakharchik05} 
and experimentally observed \cite{haller09}, that it is possible to access a regime, where the  bosonic many body system
exhibits even stronger correlations.
This opens the question, whether it is possible to enhance the correlations to a point, where the bosonic
systems forms a crystalline ground state. In this letter, we demonstrate that indeed in the presence of
an optical lattice a solid phase appears.

The transverse confinement for cold atomic gases is experimentally efficiently achieved 
using optical lattices \cite{kinoshita04,paredes04} or atomic chips \cite{hofferberth07}. Within this one-dimensional regime 
with the kinetic energy of the particles much lower than the transverse trapping frequency, 
the interaction between the particles is described by the  one-dimensional scattering 
length $a_{\rs 1D}$ \cite{olshanii98}.  Remarkably, the system can undergo a confinement 
induced resonance,  where the scattering length crosses zero.  For $a_{\rs 1D}<0$, the properties of the system
have been studied in terms of the exactly solvable Lieb-Liniger model 
\cite{lieb63.1,lieb63.2}, while at $a_{\rs 1D}=0$  the system is denoted as  Tonks-Girardeau gas.
Crossing the confinement induced resonance 
with $a_{\rs 1D}>0$ the mathematical model describing the system admits a two-particle 
bound state. Then,  the physical  state smoothly connected to the Tonks-Girardeau gas corresponds to
an highly excited state of the mathematical model; a regime denoted as
Super-Tonks-Girardeau gas \cite{girardeau10}. 

\begin{figure}[ht]
 \includegraphics[width= 1\columnwidth]{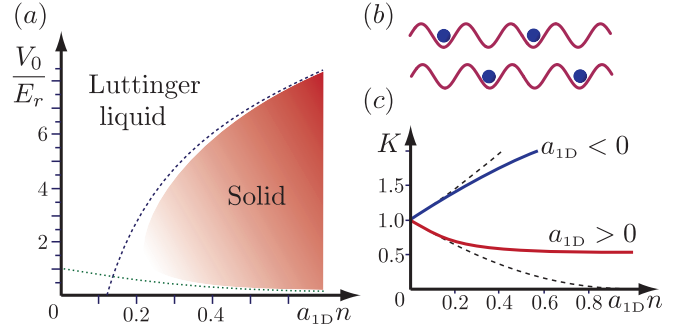}
 \caption{(a) Phase diagram: The solid phase appears at intermediate strength of the optical lattice for $a_{\rs 1D} n \gtrsim 0.2$. 
The blue (dashed) line derives from the transition within the Hubbard model, while the green (dotted) line 
describes the lower bound for the solid phase predicted from the sine-Gordon model.
 (b) Illustration for the two degenerate ground states with an atom on every second lattice site. 
 (c) Luttinger parameter derived from the exact Bethe ansatz equation.
 The dashed line denotes the asymptotic behavior $K = (1- n a_{\rs 1D})^2$. 
 }  \label{fig1}
\end{figure}

In this letter, we analyze the phase diagram within this regime and demonstrate the appearance
of a solid phase in the presence of an optical lattice with a bosonic particle on every second lattice site.
A simplified picture of this transition is that the particles acquire a finite range interaction $\sim a_{\rs 1D}$ 
and essentially behave as hard spheres \cite{girardeau10}. Then, it is natural to expect the 
appearance of a solid phase for a  density comparable to the range of 
the interaction. 
The rigorous derivation of the phase diagram follows in two steps:
First, we analyze  whether an arbitrary weak optical lattice allows to pin
the solid structure. Using the mapping to the exactly solvable sine-Gordon model,  we find, that  a finite strength of the 
optical lattice is required. Therefore, we focus on deep 
optical lattices in a second step, and provide the derivation of a Hubbard model close to the confinement 
induced resonance. The combination of the two methods allows us to identify an experimentally accessible 
region, where a solid phase appears, see Fig.~\ref{fig1}.

We start with the many-body theory describing bosonic particles confined into one-dimension.
Introducing the bosonic field operators $\psi^{\dag}(x)$ and $\psi(x)$, 
the Hamiltonian takes the form
%
%
%
\begin{eqnarray}
  H_{\rs B} & = &   \int_{\infty}^{\infty} dx \; \psi^{\dag}(x) \left[-\frac{\hbar^2}{2 m} \Delta + V(x)\right] \psi(x) \label{bosonichamiltonian}\\ 
   & & +\frac{1}{2} \int_{\infty}^{\infty} \! dx dy \:  U_{\rs B}(x-y) \psi^{\dag}(x) \psi^{\dag}(y)  \psi(y) \psi(x) \nonumber .
\end{eqnarray}
%
Here, $V(x) = V_{0} \cos^2(x k)$ accounts for the optical lattice along the tubes. 
The interaction potential between the bosons confined into the lowest state of
the transverse  trapping potential reduces to $U_{\rs B}(x) = g_{\rs B}\delta(x)$ with the
coupling strength $g_{\rs B} = - 2 \hbar^2/(m a_{\rs 1D})$ \cite{olshanii98}.
%
%
Here, the one-dimensional scattering length $a_{\rs 1D}= - a_{\perp}^2/a_{s}
\left(1- C a_{s}/a_{\perp}\right)$ is related to the three-dimensional $s$-wave
scatterling length $a_{s}$ and the transverse confining length $a_{\perp}$ with 
$C\approx 1.46$ \cite{olshanii98}.  The system exhibits a
confinement induced resonance at  $a_{s} = a_{\perp} /C$, where the coupling
strength diverges and eventually changes its character
from repulsive to attractive.

A physical interpretation of the confinement induced resonances is provided by the following property:
The 1D scattering length $a_{\rs 1D}$ describes the distance, where the scattering 
wave function for two particle crosses zero. While for $a_{\rs 1D}<0$, the zero appears in 
the unphysical region $|x|<0$, the scattering wave function exhibits a node for $a_{\rs 1D}>0$.  
This behavior is achieved by an attractive 
interaction potential $U_{\rs B}(x)$ giving rise to a bound state. 
Then, the scattering wave function is orthogonal to the bound state and consequently exhibits a node.
Note, that the sudden appearance of a bound state is an artifact of the mathematical model 
Eq.~(\ref{bosonichamiltonian}), which is  only valid in low  energy sector with the
relevant momenta $q$ satisfying the condition $q a_{\perp} \ll 1$.  In
the  physical system a bound state is always present; its position across
the confinement induced resonances has been studied in detail
\cite{bergeman03}. As a consequence, the atomic system is for all values of
$a_{\rs 1D}$ a highly excited metastable state, and losses via three-body
recombination reduce the life time of the atomic gas. This indicates that the transition
from the regime with repulsive interaction into the Super-Tonks-Girardeau gas is described by  a smooth cross-over.

In the following, 
we first focus on the limit of a very weak optical lattice $V_{0} \ll E_{r}$. Then,
the low energy properties of the strongly interacting bosonic  system are well
described within the hydrodynamics description with the bosonic field operator
$\psi(x) \sim \sqrt{n+\partial_{x} \theta}/\pi$ expressed in terms of the
long-wavelength density and phase fields $\theta(x)$ and $\phi(x)$.  The fields
satisfy the standard commutation relation $\left[\partial_{x}\theta(x),
\phi(y)\right] = i \pi \delta(x-y)$.  The effective Hamiltonian in absence of an
optical lattice reduces to
\begin{equation}
 H_{0} =  \frac{\hbar v_{s}}{ \pi} \int_{\infty}^{\infty} d x  \left[ \frac{K}{2}  \left(\partial_{x} \phi\right)^2 + \frac{1}{2 K}  \left(\partial_{x} \theta \right)^2 \right] .
 \label{lowenergyhamiltonian}
\end{equation}
The dimensionless Luttinger parameter in the strongly interacting regime
$\gamma_{\rs B} \equiv g_{\rs B} m  /  n \hbar^2 \gg 1$ reduces to $K= (1- n a_{\rs
1D})^2$ \cite{lieb63.1}. This expression remains valid in the strongly repulsive
situation with $a_{s} < 0$, as well as in the attractive case $a_{s}>0$  for $|n a_{\rs 1D}| \ll 1$ \cite{batchelor05,chen10}. In the
latter case, the dimensionless parameter $K<1$ reduces below the
non-interacting Fermi limit ($K=1$). Usually this regime can only be reached
for bosonic particles through an interaction potential with a finite range. Here,
such a finite range is achieved from the potential $U_{\rs B}(x)$ by the presence of a 
bound state and the associated node in the two-particle scattering wave function.
The behavior of the Luttinger parameter $K$ for larger 1D scattering lengths can be derived
from the exact Bethe Ansatz equation \cite{chen10}, see Fig.~\ref{fig1}. 
Note, that this behavior is in disagreement with variational Monte-Carlo simulations  predicting 
an instability for $n a_{\rs 1D} \gtrsim 0.38$ and a minimal value $K \approx 0.85$ \cite{astrakharchik05}.


Within this hydrodynamic description the weak optical lattice is a relevant perturbation 
at commensurate fillings. Here, we are interested in densities $n= 1/( m a)$ with $a= \pi/k$ the
lattice spacing and $m \in {\bf N}$ an integer. Then the Hamiltonian accounting for the optical lattice
$V_{0} \cos(k x)$ takes the form \cite{haldane81,buechler03}
\begin{equation}
 H_{\rs lattice} = u  \int dx  \cos\left(2 m \theta\right)
\end{equation}
with $u =   K  V_{0}/E_{r} ( \tilde{a}/2 a)^2$ and $\tilde{a}$ a short distance cut-off
(the  cut-off is in the range of the interparticle distance $\tilde{a}\approx 1 /n$).
The low energy description of the 
interacting bosonic system $H_{\rs eff} = H_{0}+ H_{\rs lattice}$ reduces to the quantum sine-Gordon 
model. 
%
%
%
This model is exactly solvable and exhibits a quantum phase transition from a 
gapless phase with algebraic decay in the superfluid correlation function 
$\langle  \psi^{\dag}(x) \psi(0)\rangle \sim x^{- 1/2 \tilde{K}}$ as well
as in the solid correlation $\langle n(x) n(0) \rangle \sim \cos(2 \pi n x)/x^{2 \tilde{K}}$,
to a gapped and incompressible insulator with long range order $\langle n(x) n(0)\rangle-n^2 \sim \cos(2 \pi n x)$.
Below the critical value $K<K_{m}= 2/m^2$, the transition appears for arbitrary strength of the 
lattice potential, while for a fixed value of $u$, the transition appears at the universal value 
$\tilde{K}=2/m^2$. Here, $\tilde{K}$ denotes the renormalized Luttinger parameter due to the
optical lattice; for weak optical lattices it is related to the microscopic value $K$ via the Kosterlitz-Thouless 
renormalization group flow (see \cite{kehrein99} for a review).
For a  bosonic density equal to the lattice spacing, i.e.,  $n = k/\pi$ with $m=1$, 
the phase transition takes place from the superfluid to the Mott insulating phase and has been 
previously discussed \cite{buechler03}.

In the regime with a positive 1D scattering length $a_{\rs 1D}>0$, it is now possible to access
values $K<1$. This opens the question, whether it is possible to reach the second instability
with $m=2$ and particle density $n = k/2\pi$, i.e., on average there is 
one bosonic particles distributed over two lattice sites. Then, the phase transition takes place 
from a Luttinger liquid with algebraic correlations to a crystalline phase. In addition to an excitation gap and the incompressibility,  
the crystalline phase is characterized by a long range order with a bosonic particle localized in every 
second lattice site. The ground state breaks the discrete translation invariance of the system and is 
two-fold degenerate. This property distinguishes the solid phase from the  Mott insulator at integer fillings.

The criticial value of the Luttinger parameter, where an arbitrary weak optical lattice allows to pin the bosonic 
crystalline structure reduces to $K_{2}= 1/2$. As discussed above, this regime can not be accessed. However, 
the optical lattice increases the correlations between the bosonic particles.  Using the 
Kosterlitz-Thouless renormalization group flow to lowest order in $u$ for the transition line, i.e.,  
$K = (1+u)/2$, we can expect the phase transition into the solid phase for a finite strength of the optical lattice, 
see Fig.~\ref{fig1}.  For  values of the optical lattice $V_{0 }\sim E_{r}$, the effective low energy 
theory Eq.~(\ref{lowenergyhamiltonian}) is no longer valid, and different approach is required 
for analyzing the appearance of the solid phase.


In the regime of a strong optical lattice $V_{0}> E_{r}$, the suitable  approach is to map the 
bosonic system to a Hubbard model and analyze the transition within this regime. In the 
strongly correlated regime with $\gamma_{\rs B} \gg  1$ the conventional derivation of the Hubbard model fails. Here, we first
apply an exact mapping of the strongly interacting bosonic system onto a weakly interacting Fermi gas; in the latter situation,
the conventional derivation of the Hubbard model is valid and allows us to derive the phase diagram for strong optical lattices.

This duality transformation of the strongly interacting bosons onto weakly interacting fermions has been pioneered in the past \cite{girardeau04,granger04}.
On the two particle level, it requires that the scattering wave function $\psi_{\rs B}(x)$ between two bosons with the interaction 
potential $U_{\rs B}$, is described by the a fermionic scattering wave function $\psi_{\rs F}(x)$ with a novel interaction potential $U_{\rs F}$
via $\psi_{\rs B}(x)= \sgn(x) \psi_{\rs F}(x)$ (here, $x$ denotes the relative coordinate). This property is uniquely determined by the pseudo-potential
\begin{equation}
\langle \psi | U_{\rs F} |\phi\rangle = \lim_{\epsilon \rightarrow 0^{+}} \frac{g_{\rs F}}{4}  \big[\psi'(\epsilon)+ \psi'(\!-\! \epsilon)\big]^{*} \big[\phi'(\epsilon)+\phi'(\!-\!\epsilon)\big]  
\end{equation}
with $  g_{\rs F} = 2 \hbar^2 a_{\rs 1D}/m$
the coupling strength and $\psi'=\partial_{x} \psi$  ($\phi'=\partial_{x} \phi$) the derivatives of the wave function. 
It is important to note that the role of the 1D scattering length  $a_{\rs 1D}$ is reversed in 
fermionic pseudo-potential $U_{\rs F}$ as compared to the bosonic one $U_{\rs B}$. 
As a consequence, this  mapping allows us to transform a strongly interacting bosonic model onto a 
weakly interacting Fermi system. Note, that the $\lim_{\epsilon\rightarrow 0^{+}}$ is required in order to avoid 
a ultraviolet divergence when applying the interaction potential on the Greens function. This behavior is in analogy to the
well known regulariztion of the pseudo-potential for 3D $s$-wave scattering.

\begin{figure}[ht]
 \includegraphics[width= 0.8\columnwidth]{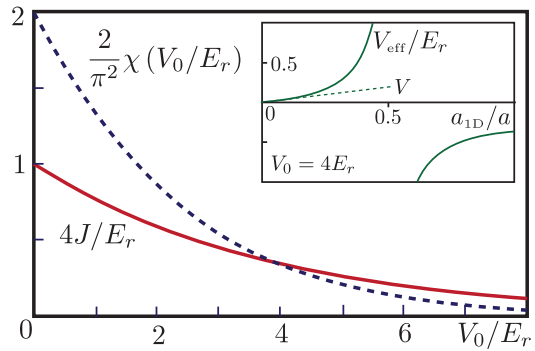}
 \caption{Tunneling amplitude $4 J$ (red) and the Wannier function overlap
$\chi$ (blue) for different strengths of the optical lattice. The inset shows the renormalization 
 of the nearest-neighbor interaction $V_{\rs eff}$  for large 1D scattering lengths accounting for the influence of higher bands and the proper  treatment of pseudo-potential $U_{\rs F}$ the 
 at $V_{0}=4 E_{r}$.}   \label{fig2}
\end{figure}

Extending this two-particle analysis to the many-body system, therefore maps the bosonic 
Hamiltonian in Eq.(\ref{bosonichamiltonian}) onto a fermionic model
\begin{eqnarray}
  H_{\rs F} & = &   \int_{\infty}^{\infty} dx \; \psi^{\dag}_{\rs F}(x) \left[-\frac{\hbar^2}{2 m} \Delta + V(x)\right] \psi_{\rs F}(x) \\ 
   & & +\frac{1}{2} \int_{\infty}^{\infty} \! dx dy \:  U_{\rs F}(x-y) \psi^{\dag}_{\rs F}(x) \psi_{\rs F}^{\dag}(y)  \psi_{\rs F}(y) \psi_{\rs F}(x) \nonumber
   \label{fermionichamiltonian}
\end{eqnarray}
with the fermionic field operators $\psi_{\rs F}^{\dag}$ and $\psi_{\rs F}(x)$. The parameter $\gamma_{\rs F}$ 
characterizing the strength of the interaction in the fermionic model is given by the ratio between 
the kinetic energy $E_{\rs kin} = \hbar^2 n^2/ m$ and the interaction energy $E_{\rs int} = n^3 g_{\rs F}$, 
i.e. $\gamma_{\rs F} =E_{\rs int}/E_{\rs kin} = 2 n a_{\rs 1D} = - 1/\gamma_{\rs B} $.
The ground state wave function $|g_{\rs F}\rangle$ of the fermionic problem is related to the ground 
state of the bosonic problem $|g_{\rs B}\rangle$, 
\begin{equation}
   \langle x_{1} , \ldots ,x_{N} | g_{\rs B}\rangle = A(x_{1},  \ldots,  x_{N})  \langle x_{1}, \ldots, x_{N} | g_{\rs F} \rangle
\end{equation}
with the total asymmetric factor $A(x_{1}, \dots, x_{N})$. For  bosons with $a_{\rs 1D}=0$, this mapping reduces to the
well known relationship between impenetrable bosons and fermions in 1D \cite{girardeau60}. However, here the situation is generalized
to arbitrary strength of the interaction potential. 

In the interesting regime with strong interactions between the bosons  $|\gamma_{\rs B} | = |1/\gamma_{\rs F}|\gg 1$, the fermionic system is weakly interacting
and the conventional approach to derive the Hubbard model is valid. Therefore, we obtain for a deep optical lattice the fermionic Hubbard model
\begin{equation}
  H_{\rs HM} = -  J \sum_{\langle  i j\rangle} c^{\dag}_{i} c_{j} + \frac{V}{2} \sum_{\langle  i j \rangle} c^{\dag}_{i} c^{\dag}_{j} c_{j} c_{i}, \label{HubbardModel}
\end{equation}
with the fermionic creation (anihilation) operator $c^{\dag}$ ($c_{i}$). In addition,
the hopping amplitude $J$ accounts for  the single particle band structure $\epsilon_{k} = -2 J \cos{k a}$, while the
fermionic pseudo-potential $U_{\rs F}$ gives rise to a dominant nearest-neighbor interaction 
\begin{equation}
   V= \frac{2}{\pi^2} E_{r} \frac{a_{\rs 1D}}{a} \:  \chi\left(\frac{V_{0}}{E_{r}}\right).
\end{equation}
Here, $\chi$ is determined by the overlap between the Wannier functions $w(x)$ on neighboring lattice sites,
\begin{displaymath}
\chi\left(\frac{V}{E_{r}}\right) =  a^3 \int \!\!  dx \left| \partial_{x} w(x) w(x\!-\!a)-  w(x) \partial_{x}w(x\!-\!a)\right|^2.
\end{displaymath}
The hopping amplitude $J$ as well as the dimensionless overlap $\chi$ can be efficiently 
determined numerically for different strengths of the optical lattice, see Fig.~\ref{fig2}. Note, that additional 
interaction terms are strongly suppressed due to the fast decay of the wannier functions.

At half filling with one particle on every second lattice site, the Hubbard model Eq.~(\ref{HubbardModel}) exhibits 
a quantum phase transition from a phase with algebraic correlations between the fermions for $J \gg V$ 
to a charge density wave with an excitation gap for $V \gg J$.  The latter phase corresponds to the interesting 
crystalline phase. The critical point for the phase transition is determined by the special point at $J=V/2$, where the 
system becomes $SU(2)$ invariant and maps to the spin-$1/2$ Heisenberg model. It is this enhanced symmetry, which fixes
the transition point to $J=V/2$ even in the one-dimensional situation. 

From the behavior of $V$ and $J$ for different strengths of the optical lattice, we can now derive 
the complete phase diagram, see Fig.~\ref{fig1}:  for very deep optical lattices the nearest neighbor interaction 
is strongly suppressed compared to the hopping term, see Fig.~\ref{fig2}, and consequently, the ground 
state is determined by a Luttinger liquid phase with algebraic correlations.  Reducing the strength of the 
optical lattice, the nearest-neighbor interaction increases and a phase transition into the solid 
phase takes place for sufficiently strong interaction $a_{\rs 1D} n \gtrsim 0.2$. For even weaker optical
lattices, the mapping to the Hubbard model breaks down, and the effective theory is given by the 
sine-Gordon model. The sine-Gordon model requires a finite strength of the optical lattice for the 
appearance of the solid phase. Therefore, a second phase transition
takes place for decreasing optical lattice, and the system enters again the Luttinger liquid phase, i.e., the system exhibits a remarkable  reentrant feature. Consequently, we 
predict the existence of a solid phase for cold atomic gases at strong interactions $a_{\rs 1D} n \gtrsim 0.2$
and intermediate  the optical lattices $V\approx 3 E_{r}$.

Finally, we have to verify the validity of the Hubbard model in the interesting regime with $n a_{\rs 1D} \gtrsim 0.2$. 
The derivation of the Hubbard model involves two approximations:  (i) first, we restrict the analysis onto the lowest 
Bloch band, i.e., we introduce a high energy cut-off $\Lambda \gtrsim a$ determined by the lattice spacing. 
(ii) Second, the interaction potential $U_{\rs F}$ is treated without the proper regularization.
The influence of these two-approximation has recently been studied in detail for the derivation of the Hubbard model 
in a three-dimenionsal optical lattice \cite{buechler10}. Here, the 
situation is equivalent and the main results can be directly carried over. It follows, that the Hubbard model
is correct for weak interactions $a_{\rs 1D} \ll a$, while in the interesting parameter range $a_{\rs 1D} n \sim 0.2$ corrections 
from higher bands and the proper treatment of the interaction potential appear. 
The main influence is a renormalization of the nearest neighbor interaction strength, which takes the from $ V_{\rs eff} = V/(1+ \eta V/E_{r})$.
%
%
%
Here, $\eta$ describes a dimensionless parameter which in general derives from a full numerical analysis.
However, due to the duality mapping between the Bosons and Fermions, we know that in the limit $a_{\rs 1D}/a \rightarrow \infty$
the system has to reproduce the scattering of non-interacting bosons. This  condition fixes the parameter to $\eta = - E_{r}/2J$.
Therefore, we find that the influence of higher bands and the proper treatment of the interaction potential increases the strength of the
nearest-neighbor interaction, see Fig.~\ref{fig2}. Therefore, we  expect that the solid phase appears even for weaker interactions
than  shown in Fig.~\ref{fig1}.

The experimental setup required for the observation of the solid phase can be achieved by the combination of strong transverse 
confining by an optical lattice with a Feshbach resonance to tune the strength of the $s$-wave scattering length. 
Such a setup has recently been realized for the observation of correlations beyond the Tonks-Girardeau regime \cite{haller09}.
An additional weak optical lattice along the tubes then opens the path to the experimental search of the solid phase.
Finally, it is important to note, that the behavior of losses by crossing the confinement induces resonance are not yet well understood.
However, 
for increasing 1D scattering length, additional terms to the Hamiltonian breaking the integrability of the model, 
e.g.,   corrections from higher transverse  states and additional non-universal three-body interactions, 
provide a decay rate and eventually an instability of the Super-Tonks-Girardeau gas 
towards the formation of  bound states; such a behavior was observed within the variational Monte Carlo simulations \cite{astrakharchik05}.
This implies a finite lifetime for the realization of the experiments and suggests that the search for  the solid phase should be performed for intermediate interaction strengths
 $n a_{\rs 1D} \sim 0.4$. 

We thank M. Girardeau for helpful discussions. The work was supported by the 
Deutsche Forschungsgemeinschaft (DFG) within SFB/TRR 21 and National Science Foundation under Grant No. NSF PHY05-51164.



\begin{thebibliography}{10}

\bibitem{girardeau60}
M.~Girardeau,
\newblock J.\ Math.\ Phys. {\bf 1}, 516 (1960).

\bibitem{kinoshita04}
T.~Kinoshita, T.~R. Wenger, and D.~S. Weiss,
\newblock Science {\bf 305}, 1125 (2004).

\bibitem{haller10}
E.~Haller et~al.,
\newblock Nature {\bf 466}, 597 (2010).

\bibitem{buechler03}
H.~P. B{\"u}chler, G.~Blatter, and W.~Zwerger,
\newblock Phys.\ Rev.\ Lett. {\bf 90}, 130401 (2003).



\bibitem{astrakharchik05}
G.~E. Astrakharchik et~al., 
\newblock Phys. Rev. Lett. {\bf 95}, 190407 (2005).

\bibitem{haller09}
E.~Haller et~al.,
\newblock Science {\bf 325}, 1224 (2009).

\bibitem{paredes04}
B.~Paredes et~al.,
\newblock Nature {\bf 429}, 277 (2004).

\bibitem{hofferberth07}
S.~Hofferberth et~al.,
\newblock Nature {\bf 449}, 324 (2007).

\bibitem{olshanii98}
M.~Olshanii,
\newblock Phys.\ Rev.\ Lett. {\bf 81}, 938 (1998).

\bibitem{lieb63.1}
E.~H. Lieb and W.~Liniger,
\newblock Phys.\ Rev. {\bf 130}, 1605 (1963).

\bibitem{lieb63.2}
E.~H. Lieb,
\newblock Phys.\ Rev. {\bf 130}, 1616 (1963).

\bibitem{girardeau10}
M.~D. Girardeau and G.~E. Astrakharchik,
\newblock Phys. Rev. A {\bf 81}, 061601(R) (2010).

\bibitem{bergeman03}
T.~Bergeman, M.~G. Moore, and M.~Olshanii,
\newblock Phys. Rev. Lett. {\bf 91}, 163201 (2003).

\bibitem{batchelor05}
M.~T. Batchelor et~al.,  
\newblock Stat. Mech. {\bf 2005}, L10001 (2005).

\bibitem{chen10}
S.~Chen et~al., 
\newblock Phys. Rev. A {\bf 81}, 031609 (2010).

\bibitem{haldane81}
F.~D.~M. Haldane,
\newblock Phys.\ Rev.\ Lett. {\bf 47}, 1840 (1981).

\bibitem{kehrein99}
S.~Kehrein,
\newblock Phys.\ Rev.\ Lett. {\bf 83}, 4914 (1999).

\bibitem{girardeau04}
M.~Girardeau, H.~Nguyen, and M.~Olshanii,
\newblock Opt. Comm. {\bf 243}, 3  (2004).

\bibitem{granger04}
B.~E. Granger and D.~Blume,
\newblock Phys. Rev. Lett. {\bf 92}, 133202 (2004).

\bibitem{buechler10}
H.~P. B\"uchler,
\newblock Phys. Rev. Lett. {\bf 104}, 090402 (2010).



\end{thebibliography}

\end{document}